\pdfoutput=1

\documentclass[sigconf]{acmart}

\usepackage{booktabs} % For formal tables
\usepackage{graphicx}
\usepackage{subfigure}
\usepackage{CJKutf8} 
\setcopyright{rightsretained}
\usepackage{hyperref}
\usepackage{color}
\hypersetup{colorlinks,
  urlcolor=[rgb]{0.5,0.5,0.5},
            citecolor=blue,
            linkcolor=blue}
\usepackage{threeparttable}

\acmISBN{} 
\acmDOI{}

\acmConference[LBRS@RecSys'18]{the Late-Breaking Results track part of the Twelfth ACM Conference on Recommender Systems}{October 2--7, 2018}{Vancouver, BC, Canada}
\acmBooktitle{Proceedings of the Late-Breaking Results track part of the Twelfth ACM Conference on Recommender Systems (RecSys '18), Vancouver, BC, Cananda, October 2--7, 2018}
\acmYear{2018}
\copyrightyear{2018}

\begin{document}
\title{Superhighway: Bypass Data Sparsity in Cross-Domain CF}
%\titlenote{Produces the permission block, and
%  copyright information}
%\subtitlenote{The full version of the author's guide is available as
%  \texttt{acmart.pdf} document}

\author{Kwei-Herng Lai}
\authornote{Both authors contributed equally to the paper.}
\affiliation{%
  \institution{Academia Sinica, Taiwan}
}
\email{khlai@citi.sinica.edu.tw}

\author{Ting-Hsiang Wang}
\authornotemark[1]
\affiliation{%
  \institution{Academia Sinica, Taiwan}
}
\email{thwang@citi.sinica.edu.tw}

\author{Heng-Yu Chi}
\affiliation{%
  \institution{KKBOX Inc., Taiwan}
}
\email{henrychi@kkbox.com}

\author{Yian Chen}
\affiliation{%
  \institution{KKBOX Inc., Taiwan}
}
\email{annchen@kkbox.com}

\author{Ming-Feng Tsai}
\affiliation{
   \institution{National Chengchi University, Taiwan}
}
\email{mftsai@nccu.edu.tw}

\author{Chuan-Ju Wang}
\affiliation{
   \institution{Academia Sinica, Taiwan}
}
\email{cjwang@citi.sinica.edu.tw}

\renewcommand{\shortauthors}{K. Lai, T. Wang, H. Chi, M. Tsai, and C. Wang}

% Sean's note spot:
%   1) Domain definition: A system is decomposed to: 1) user and 2) item domains.
%       We address both.
%   2) The distance is commutative
%   3) Joint recommendation?
%   4) Source and target embedding space amalgamation
%   5) Create user stitching from item stitching to join the domains/embeddings spaces.
%   6) a user-user express path, with respect to the strength of the \emph(highway),
%   7) a user-item-user \emph{bridge} path.
%   8) Our method can be easilly extend to incorporate more source domains
%   9) (Future) Joint recommendation
%   10) We are solving a RS problem with RS (Recommend users to users via CF)

% Reference:
%   1) Cross-Domain Collaborative Filtering: A Brief Survey
%   2) LINE
%   3) DeepWalk
%   4) LibFM

\begin{abstract}
    % Write about merits,
    %   1) Solve sparsity and filtering problem
    %   2) Provide NARWal to merge embedding space via point-wise anchor instead
    %       of window
    %   3) Improves in both target and source domain.
    %   4) Particularly good for RW-based methods.
    % Write about the framework which solve the problem,
    %   1) Superhighway construction phase:
    %       a) Bridge point identification: Critical capacity rate, alpha
    %       b) Relation scaling: Relation scaling rate: beta
    %   2) Embedding learning phase:
    %       a) NARWal instead of window-based.
    % Write about how proposed method address the problem.
    %   1) By superhighway construction, which aligns different but related
    %       embedding spaces
    %   2) By node-anchoring instead of trail-window anchoring

    Cross-domain collaborative filtering (CF) aims to alleviate data sparsity
    in single-domain CF by leveraging knowledge transferred from related
    domains.
    Many traditional methods focus on enriching compared neighborhood
    relations in CF directly to address the sparsity problem.
    % However, beyond the existence of transferable knowledge, many existing
    % methods assume additional conditions to facilitate the transfer,
    % e.g., source domain denser than target domain and prior knowledge about
    % the correlation between overlapped entities.
    In this paper, we propose \emph{superhighway} construction, an alternative
    explicit relation-enrichment procedure to improve recommendations by
    enhancing cross-domain connectivity.
    Specifically, assuming partially overlapped items (users), superhighway
    bypasses multi-hop inter-domain paths between cross-domain users (items,
    respectively) with direct paths to enrich the cross-domain connectivity.
    The experiments conducted on a real-world cross-region music dataset and a
    cross-platform movie dataset show that the proposed superhighway
    construction significantly improves recommendation performance in both
    target and source domains.

\end{abstract}

\keywords{recommendation; data sparsity; cross-domain; knowledge transfer}

\maketitle

\section{Introduction}
    % Cross-domain RS adaptation/localization is a fundamental problem for expanding e-commerce
    %       1) when entering foreign market, like spotify in Japan
    %       2) when acquire and merge compeitior, Walmart acquires Jet.com
    
    % Can be though of in CF . (but CF isn't the only RS solution, what makes CF
    %       particularly important here.)  Why is CF worthy particularly mentioning.
    
    % We can assume knowledge about embeddings is given???
    
    % Traditionally, cross-domain problem is solved via MF, there are two approaches:
    %       1) Joint MF: matrix line up entities in both domains
    %       2) Transfer learning: two-stage MF
    % How is MF-based CS conducted traditioinally????
    % Here we leverage the flexibility of graph embedding to take more fine-grained contexts:
    %       1) DeepWalk
    %       2) NARWAl
    %       3) LINE: resemble MF
    % We can think of localization as achieving the merging of the source
    %       and target space rationalized using user-preference pattern 
    % Traditional methods require overlap of user or user alignment, which is
    %       difficult.
    
    % title:Superhighway Construction: Bypass Data Sparsity in Cross-Domain Collaborative Filtering
    % tabular/observed VS structured/inferenced

    Collaborative filtering (CF) in recommender systems is highly susceptible
    to data sparsity as the method analyzes observed user-item interactions
    solely.
    In modern e-commerce, as the number of items and users skyrockets and
    dwarfs the growth of user-item ratings in comparison, data sparsity takes
    an increasing toll on the performance of CF-based recommender systems.
    In response to such a vital issue, cross-domain CF is proposed to enhance
    recommendation quality in a given target domain by leveraging knowledge
    transferred from related source domains.

    As data sparsity in single-domain CF remarks the lack of observed rating
    data, intuition suggests to alleviate the sparsity issue via explicitly
    populating relations in a cross-domain system.
    In the literature, many traditional methods have been proposed to directly
    enrich the compared neighborhood in CF, which, for example, attach
    additional intra-domain edges in target domains~\cite{li2009can} or
    inter-domain edges in overlapped regions~\cite{cremonesi2011cross}.
    However, such methods typically require additional assumptions; for
    example, the source domain has to be denser than the target
    domain~\cite{li2009can}.
    %and the existence of the correlation matrix for
    %overlapped entities~\cite{cremonesi2011cross}.
    
\begin{figure}[t]
  \includegraphics[height=4.5cm]{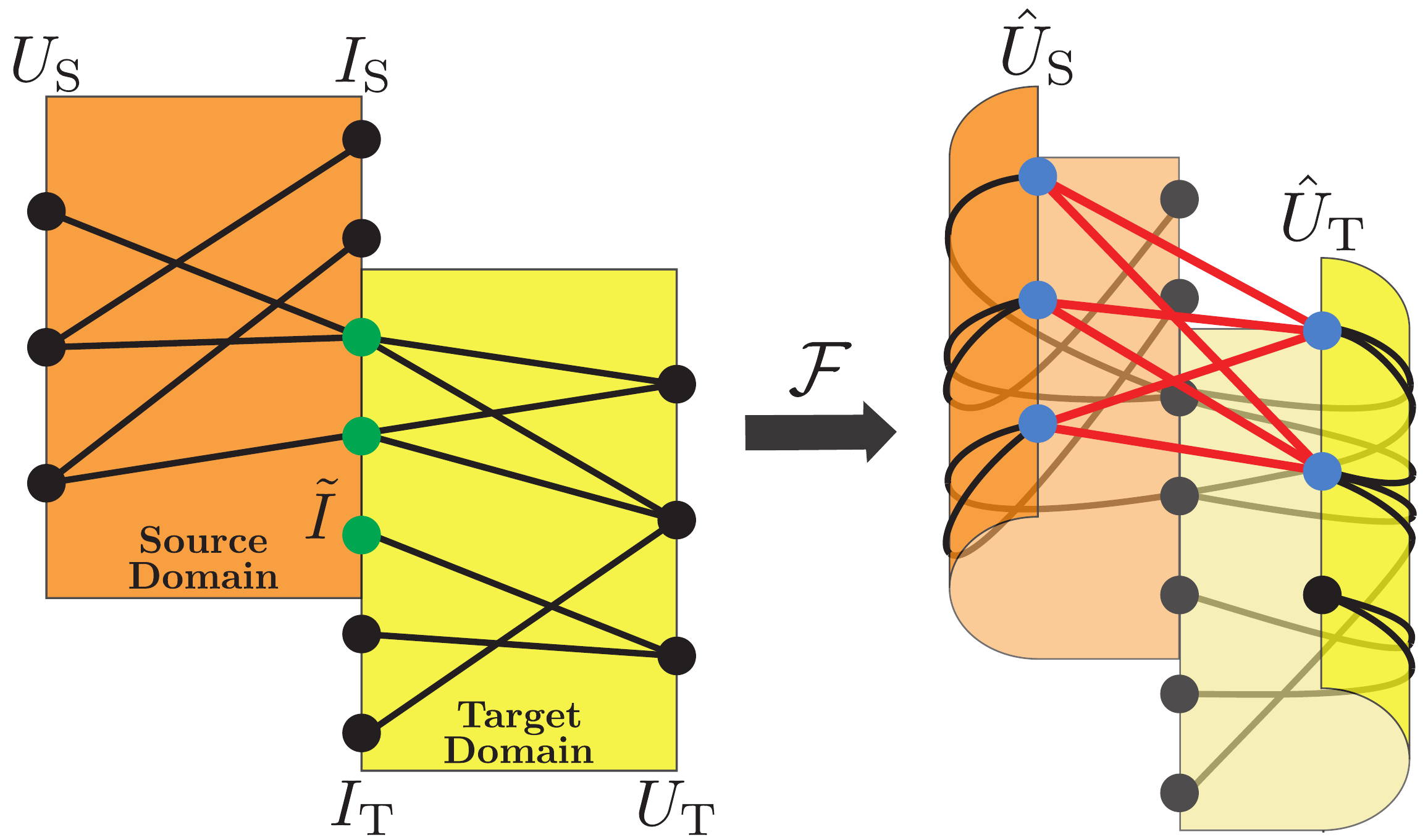}
    \vspace{-0.3cm}
    \caption{Illustrative example for superhighways}
  \label{fig:fig1}
\end{figure}

    In this paper, our superhighway construction establishes a new type of
    relations by means of inference based on existing relations, which allows
    the source and the target domains to mutually improve due to the enhanced
    cross-domain connectivity.
    % In the experiments, we demonstrate the effectiveness of our
    % connectivity-boosting approach in improving CF performance and thus
    % bypass the data sparsity problem.
    % Specifically, our tests are conducted on single domain, double domains,
    % and double domains with superhighway using both the traditional matrix
    % factorization method and the new graph-based methods with random walk.
    The construction of superhighways consists of two steps: 1) the
    identification of cross-domain user candidates suitable for superhighway
    construction, and 2) weight scaling for superhighways to optimize
    cross-domain space alignment.
    Figure~\ref{fig:fig1} illustrates the connectivity enhancement brought
    forth by superhighways (red bold lines), which provide additional leverage
    of combining the source and the target domains.

\section{Methodology}
    In collaborative filtering (CF), user-item interactions are commonly
    captured using a bi-adjacency matrix $M=(m_{ij})\in \mathbb{R}^{|U| \times
    |I|}$, where $U$ and $I$ denote the sets of users and items, respectively;
    $m_{ij}=1$ if there exists an observed association for user $i$ and item
    $j$, and otherwise, $m_{ij}=0$.
    The matrix $M$ can also be represented as a bipartite graph $G=(U,I,R)$,
    where $R=\{(i,j)\,|\,m_{ij}=1\}$.

    Given a cross-domain system with source domain
    $G_{\text{S}}=(U_{\text{S}},I_{\text{S}},R_{\text{S}})$ and target domain
    $G_{\text{T}}=(U_{\text{T}},I_{\text{T}},R_{\text{T}})$ such that the set
    of shared items $\tilde{I} = I_{\text{S}} \cap I_{\text{T}} \neq
    \varnothing$, a \emph{highway} is defined as a path between user $u_i\in
    U_{\text{S}}$ and $u_j\in U_{\text{T}}$ through shared items in
    $\tilde{I}$.
    To enrich the cross-domain connectivity, the \emph{superhighway
    construction}, denoted as an operation $\mathcal{F}$, establishes direct
    relations between candidate users $u_i\in \hat{U}_{\text{S}}$ and $u_j\in
    \hat{U}_{\text{T}}$, where $\hat{U}_{\text{S}}\subseteq {U}_{\text{S}}$ and
    $\hat{U}_{\text{T}}\subseteq {U}_{\text{T}}$ are the sets of
    candidate users from the source and target domains,
    respectively.
    Consequently, the new graph $\mathcal{F}(G_{\text{S}},G_{\text{T}})$, which
    is more connected than the naively joined graph $G_{\text{S}}\cup
    G_{\text{T}}$, can then be used for CF.
    The candidate user sets $\hat{U}_{\text{S}}$ and $\hat{U}_{\text{T}}$
    mentioned above are defined as 
	\begin{equation}
        \hat{U}_{d} = \left\{u\,\left|\, u \in U_{d}, \frac{|\mathcal{N}(u) \cap \tilde{I}|}{|\mathcal{N}(u)|}\right.\geq\alpha\right\},
      \label{eq:candidate}
	\end{equation}
    where $d \in \{\text{S}, \text{T}\}$, $\mathcal{N}(u)$ is the set of
    neighbors of $u$, and~$\alpha$ is the predefined smoothness
    threshold.
    %Note that in this way, we can utilize the shared items in $\tilde{I}$ as a
    %sub-domain to control the smoothness between user interactions.
    
	%\subsection{Superhighway Construction}
    While many proposed cross-domain CF methods approach the data sparsity
    problem by directly enriching the compared neighborhood (e.g., the
    user-item relations in each domain), we instead
    % seek to raise recommendation performance by
    enhance cross-domain connectivity.
    Specifically, we establish superhighways between users from each of the
    candidate sets defined in Eq.\ (\ref{eq:candidate}), resulting additional
    $|\hat{U}_{\text{S}}|\times|\hat{U}_{\text{T}}|$ superhighways. 
    The weight between each pair of users is defined as
    \begin{equation}
      w = \beta \times |\mathcal{N}(u_{i}) \cap \mathcal{N}(u_{j})|, 
    \end{equation}
    where $u_{i} \in \hat{U}_{\text{S}}, u_{j} \in \hat{U}_{\text{T}}$, and
    $\beta$ is the scaling factor for the strength of domain alignment.
    Notice superhighways are kept weighted to provide fine-grained alignment
    between the source domain and the target domain.
    % The rationale behind this design is that, while the tabular neighborhood
    % comparison of traditional CF for items only utilizes user-item information,
    % the nuanced user-user relations facilitate the difference between
    % alignments in the learning process.

%	\subsection{Learning Representation}
%	        Furthermore, we propose to learn the representations of user and items for recommendation by random walk based network embedding technique, 
%                which is able to traverse the constructed network and model the proximity between the nodes on passage. Specifically, the proposed method firstly
%                samples a root node $v_{i}$ by the degree distribution of the whole network, then samples the neighborhood $N(v_{i})$ of the root node $v_{i}$ by 
%                a biased random walker which traverse the network based on the edge weight. After that, we minimize the negative log-likelihood of proximity of
%                sampled root node $v_{i}$:
%		\begin{equation}
%            O = - \sum_{v_{j}\in N(v_{i})} \log P((v_{j})|\Phi(v_{i}))
%			\label{eq:obj}
%		\end{equation}
%		where $\Phi(v_{i})$ is denoted as the representation of a sampled root node $v_{i}$, and $N(v_{i})$ represents the sampled neighborhood of root node $v_{i}$ by random walk sampling.

    \begin{table}
      \setlength{\tabcolsep}{2pt}
      \begin{threeparttable}[b]
        \caption{Data statistics\vspace{-0.4cm}}
        \label{tb:data}
        \begin{tabular}{ccccc}
          \toprule
          & \multicolumn{2}{c}{Cross-region music dataset} & \multicolumn{2}{c}{Cross-platform movie dataset} \\
          \cmidrule(lr){2-3}\cmidrule(lr){4-5}
            & KKBOX\_R1 (S)                & KKBOX\_R2 (T)    & Netflix (S)      & Movielens (T)   \\
          \midrule
          User	     & 184,607              & 72,042      & 480,189       & 69,878       \\ 
          Item       & 529,457              & 87,889      & 17,779        & 10,677       \\
          Rating     & 21,961,070           & 4,473,052   & 100,480,507   & 10,000,054   \\
          \bottomrule
        \end{tabular}
        \begin{tablenotes}
          \vspace{-0.1cm}
        \item[] {$^*$S and T denote the source and target domains, respectively.}
        \end{tablenotes}
      \end{threeparttable}
    \end{table}

\section{Experiments}

    In order to validate the effectiveness of the proposed superhighway
    construction on cross-domain collaborative filtering (CF), we conducted
    query-based recommendation~\cite{HPE} and used items as queries.
    %Specifically, we leverage an item favored by user as a query to retrieve
    %items which is also favored by the user.
    %\subsection{Experiment settings}
    Our experiments employ two sets of real-world cross-domain datasets:
    1) KKBOX\_R1--KKBOX\_R2, a cross-region music dataset (R denotes region);
    2) Movielens--Netflix, a cross-platform movie dataset.
    The statistics of the datasets are listed in Table~\ref{tb:data}.
    The cross-domain datasets are organized into three structures for training:
    1) single: denoting the original target domain, $G_{\text{T}}$; 2) highway:
    denoting the naive concatenation of the source and target domains,
    $G_{\text{S}}\cup G_{\text{T}}$; and 3) superhighway: denoting the naive
    concatenation augmented with superhighways to enhance cross-domain
    connectivity, $\mathcal{F}(G_{\text{S}}\cup G_{\text{T}})$.
    With these three structures, models (i.e., user and item embedding) are
    trained using traditional matrix factorization and two network embedding
    algorithms: DeepWalk~\cite{deepwalk} and HPE~\cite{HPE}.
    In addition, transfer learning is conducted for the single structure via
    pre-training on the source domain and then fine-tuning on the target
    domain~\cite{tang2015pte}.
    % the results of which are denoted as Pretrained
    % and listed in the parentheses in Table~\ref{tb:het_exp}.
    For each algorithm, we also find the best combination of $\alpha$ and
  $\beta$ in the interval of $(0.0,1.0]$ and $[0.5,1.5]$, respectively, with
  $0.1$ increment.

    %\subsection{Cross-domain collaborative filtering}
	\begin{table}[t]
      \setlength{\tabcolsep}{3.5pt}
		\caption{Recommendation performance (MAP@10)}
                \vspace{-0.4cm}
		\centering
		\label{tb:het_exp}
		\begin{tabular}{cccccc}
        \toprule    &       &   MF              &   DeepWalk        &   HPE         \\
        \midrule    &   Single (Pretrained)  &   30.4 (30.3)     &   19.6 (22.2)     &   14.2 (27.8) \\
        Music            &   Highway			&   30.5            &   0.193	        &   0.2	        \\
                    &   Superhighway    &   32.4            &   22.6	        &   31.1        \\
		\midrule
        &   Single (Pretrained)	&   5.5 (5.3)       &   2.8 (1.7)	    &   4.2 (6.3)   \\ 
        Movie            &   Highway			&   1.4             &   2.0	            &   0.014       \\
                    &   Superhighway    &   6.8             &   4.0	            &   7.4         \\
		\bottomrule
		\end{tabular}
	\end{table}
    Table~\ref{tb:het_exp} compares the performance on the target domain of the
    above three structures.
    Note that most models perform worse when training on the highway structure
    than on the single structure; this phenomenon is likely due to the naive
    combination of the two domains actually aggravates data sparsity in the
    system, demonstrating the mere introduction of transferable knowledge is
    insufficient.
    % for improving cross-domain CF.
    In contrast, the superhighway structure reduces data sparsity and
    facilitates structural alignment between the source and the target domains
    by enhancing cross-domain connectivity, thereby creating a mutually
    enriching relationship.
    Hence, superhighway improves CF-based recommendation across all algorithms,
    making it widely applicable.
    In addition, it is worth noting that superhighway, as a user-user relation,
    does not enrich item neighborhood.
    Therefore, the improvement in matrix factorization suggests superhighways
    ``bypass'' the data sparsity problem in cross-domain CF, addressing the
    problem indirectly by enhancing the connectivity of the cross-domain
    system.
    Moreover, as traditional cross-domain improvements are often directional,
    i.e., source domains facilitate target domains, superhighway also improves
    recommendation performances on source domains; e.g., HPE improves from
    $2.1$ to $4.4$ for the music dataset and from $4.4$ to $4.6$, for the movie
    dataset.

%    \begin{table}
%        \caption{Mutual Enhancement on HPE}
%        \vspace{-0.3cm}
%        \label{tb:me}
%        \begin{tabular}{cccccc}
%	    \toprule
%            & Music  MAP@10(\%)		& Source & Target   \\
%	    \midrule
%            & Single 	                & 2.1    & 14.2     \\ 
%            & Superhighway              & 4.4    & 31.1     \\
%	    \bottomrule
%        \end{tabular}
%        \vspace{5pt}
%        \begin{tabular}{cccccc}
%	    \toprule
%            & Movie MAP@10(\%)		& Source & Target   \\
%	    \midrule
%            & Single 	                & 4.4    & 4.2     \\ 
%            & Superhighway              & 4.6    & 7.6     \\
%	    \bottomrule
%        \end{tabular}
%    \vspace{-0.5cm}
%    \end{table}

\section{Conclusions}
This paper proposes an explicit relation-enrichment procedure,
\emph{superhighway construction}, to bypass data sparsity in single-domain
collaborative filtering by enhancing cross-domain connectivity using
self-contained inference.
In our approach, superhighways are generated based on suitable (interaction
smoothness) highways and then scaled for domain space alignment.
According to the results form cross-region music dataset and cross-platform
movie dataset, the constructed superhighways not only facilitate improvements
across all tested models but also lead to improvements in the source domains,
making it widely applicable.
%@@In future work, we will study how the degree of overlap effects
%@@superhighway performance and how superhighway may facilitate cross-domain
%@@recommendation, i.e., recommend items across domains.

\bibliographystyle{ACM-Reference-Format}
\bibliography{paper} 

\end{document}